# Nano-ARPES observation of a flat band at the $\overline{\mathrm{M}}$ point in 2.4° twisted bilayer graphene

**Giovanna Feraco[1#], Przemysław Przybysz[1,2], Oreste De Luca[1$], Ying Wang[1%], Pavel Dudin[3], José Avila[3], Jianting Ye[1], Paweł Dąbrowski[2], Paweł J. Kowalczyk[2], Jagoda Sławińska[1], Petra Rudolf[1] and Antonija Grubišić-Čabo[1*]**

[1] Zernike Institute for Advanced Materials, University of Groningen, Nijenborgh 3, 9747 AG Groningen, The Netherlands
[2] Faculty of Physics and Applied Informatics, University of Łódź, Pomorska 149/153, 90–236 Łódź, Poland
[3] Synchrotron SOLEIL, Université Paris-Saclay, Saint Aubin, BP 48, 91192 Gif sur Yvette, France
[#] Present address: European X-ray Free Electron Laser GmbH, Holzkoppel 4, 22869 Schenefeld, Germany
[$] Present address: Dipartimento di Fisica, Università della Calabria, Via Pietro Bucci, 87036 Rende, Italy.
[%] Present address: Department of Biomedical Signals and Systems, University of Twente, 7500 AE Enschede, The Netherlands

*Author to whom any correspondence should be addressed.

**E-mail:** a.grubisic-cabo@rug.nl

## Abstract

The emergence of flat bands in twisted bilayer graphene at the so-called magic angle reveals a range of intriguing phases, such as superconductivity and Mott insulating states, resulting from electrons entering a correlated state. This establishes twisted bilayer graphene as an excellent platform for exploring correlation effects in many-body phases and across various twist angles. However, in most cases, the twist angle is not an easy parameter to control because it exhibits structural instabilities such as relaxation and reconstruction that reduce the correlation effects. Here, we report the unexpected appearance of a flat band at the $\overline{\mathrm{M}}$ point far from the Fermi level in a 2.4° twisted bilayer graphene, directly visualized by angle-resolved photoemission spectroscopy. Our work is the first observation of flat bands away from the $\overline{\mathrm{K}}$ point in twisted bilayer graphene and highlights the need to probe the entire Brillouin zone to fully capture electronic states in twisted structures.

## 1. Introduction

Single-layer graphene has been extensively explored over the last few decades, revealing remarkable optical, thermal, and, above all, electronic properties [1-3]. Its special electronic properties arise from a linear energy dispersion E(k) at the $\overline{\mathrm{K}}$ points of its Brillouin zone. As a consequence, graphene displays exceptional transport properties, characterized by very high carrier mobility, making it highly attractive for electronic applications [3]. The success of graphene has naturally motivated the investigation of more complex multilayer graphene-based systems, whose electronic properties can differ significantly from those of a single layer [4-6]. In addition to stacking multiple layers, the weak interlayer coupling characteristic of van der Waals materials such as graphene, allows the layers to be rotated relative to one another, introducing an additional structural degree of freedom that is generally unavailable in conventional three-dimensional crystals [7].

This relative rotation of the two layers gives rise to twisted van der Waals structures and it is closely associated with the emergence of the moiré effect [8, 9]. In particular, when one graphene layer is placed on top of another with a twist angle ϑ the interference between the two atomic lattices produces a larger-scale periodic pattern known as a moiré pattern. This moiré pattern forms a superlattice arising from the long-range modulation of the atomic density and its characteristic periodicity depends on the twist angle between the layers and for commensurate lattices can be written as: $\lambda_M = \frac{a}{2\sin(\vartheta/2)}$; where $\lambda_M$ is the wavelength of the moiré pattern and $a$ the lattice constant which amounts to 2.46 Å in graphene [8, 9]. In certain conditions, the moiré superlattice can effectively modulate the electronic properties of the material [10]. For graphene, this effect is particularly significant at a 1.1° twist angle, also called the magic angle, where twisted bilayer graphene (TBG) presents flat bands in the vicinity of the Fermi level, and the electrons are extremely localized [11]. The reduced kinetic-energy scale enhances the relative importance of electron-electron interactions, giving rise to a rich phase diagram that includes correlated insulating and superconducting states [9, 12, 13]. The remarkable range of phenomena observed in TBG, along with the relative ease of experimentally probing them, makes it an ideal platform for investigating how geometry, interlayer coupling, and electronic correlations combine to generate emergent quantum phases. This provides strong motivation for a better understanding of graphene's twisted electronic band structure. Consequently, most experimental and theoretical studies of TBG, including those using angle-resolved photoemission spectroscopy (ARPES) studies, have focused on the electronic structure near the $\bar{\mathrm{K}}$ point and at twist angles close to the magic-angle regime. ARPES, particularly in its spatially resolved nano- and micro-ARPES implementations, provides direct access to the momentum-resolved band structure and has been instrumental in visualizing moiré-induced electronic states [14]. Flat bands and hybridization gaps have been observed close to the magic angle, [15] but weakly dispersive states have also been reported at larger twist angles, including angles of approximately 3° and 12.2°, as well as in twisted monolayer–bilayer graphene structures. [16-19] A flat feature has furthermore been detected in the proximity of the $\bar{\mathrm{K}}$ point near the Fermi level for TBG with a twist angle of approximately 3.2° [20]. These observations indicate that weakly dispersive states are not necessarily restricted to the immediate vicinity of the magic angle. Nevertheless, the electronic structure of TBG at larger twist angles remains incompletely understood, particularly away from the $\bar{\mathrm{K}}$ point. This limitation partly reflects the experimental and theoretical emphasis on the low-energy Dirac bands due to their importance for transport, despite the possibility that moiré coupling, structural relaxation, strain, or interfacial interactions may generate localized states elsewhere in momentum space. For example, the $\bar{\mathrm{M}}$ point hosts van Hove singularities [5] that are relevant for optical conductivity [21], can enhance electron-phonon coupling [22] and induce superconductivity if brought near the Fermi level via doping [23, 24] and can even host emergent symmetries [25]. In TBG, the moiré superlattice may modify these states; however, the electronic structure of TBG at the $\bar{\mathrm{M}}$ point has not been experimentally probed by ARPES to date.

In this study we report the electronic structure of twisted bilayer graphene with a measured twist angle of 2.4° *via* nano-ARPES measurements along the $\bar{\Gamma} - \bar{\mathrm{M}}$ and $\bar{\Gamma} - \bar{\mathrm{K}}$ high-symmetry directions. No flat band is observed near the $\bar{\mathrm{K}}$ point. In contrast, a pronounced weakly dispersive feature is detected near the $\bar{\mathrm{M}}$ point at higher binging energy. Tight-binding and density functional theory (DFT) calculations reproduce the principal dispersive bands but do not account for this additional feature. The discrepancy suggests that its origin may involve structural or interfacial effects that are not captured by the present models. Our findings therefore extend the experimental exploration of TBG beyond the conventional focus on the low-energy electronic structure near $\bar{\mathrm{K}}$ point and motivate further investigation of moiré-derived states throughout the full Brillouin zone.

## 2. Methods

The mechanical exfoliation method was used to obtain monolayer graphene (HQ graphene) and thin *h*-BN (HQ graphene) flakes on $SiO_2$/Si (University wafer) substrates [26]. This process involves repeatedly pressing Scotch tape against the material to peel off layers until a single or few-layered flake is achieved. Once the desired thin flakes are obtained on the substrates, the dry transfer technique is employed to precisely position them onto a target substrate. This sample was fabricated using the conventional "tear and stack" dry transfer method with a rotation involved [27]. A polymer stamp, consisting of a glass slide, a polydimethylsiloxane (PDMS, Sigma Aldrich) substrate, and a poly (bisphenol A carbonate) (PC, Sigma Aldrich) film, was used to pick up one of the cut graphene pieces at 90 °C. This piece was then rotated by 2°, with the rotational transfer stage offering a resolution of 1/60°. The rotated graphene layer was carefully placed on top of the remaining half. After assembling the TBG stack, a thick *h*-BN flake was picked up at approximately 100 °C and positioned onto a 30 nm thick Ti/Au film on a marked $SiO_2$/Si wafer by melting the PC film at temperatures above 180 °C. The PC film was removed by immersing the assembly in chloroform heated at 45 °C, for 40 min.

While a gold layer provides a reliable conductive substrate for ARPES measurements to avoid charging, the use of *h*-BN was motivated by the need for an atomically flat surface in order to obtain sharp band features [28]. To avoid the formation of a significant moiré pattern with graphene, a large twist angle between graphene and *h*-BN was chosen. At sufficiently large angles, the moiré wavelength becomes very short, suppressing the electronic influence of an additional moiré pattern [29, 30]. To ensure that the sample surface was atomically clean, as necessary for ARPES measurements, atomic force microscopy (AFM) was performed after each step to check the quality of the flakes and clean the surface from impurities and bubbles trapped between the layers. The sample was also annealed for 30 min *in situ* at 150 °C right before the ARPES measurement to desorb water and hydrocarbons that accumulated during the transfer through air to the beamline [18].

A conductive-mode AFM (c-AFM) system, operating in contact mode (Dimension, Bruker), was employed to simultaneously map topography (height channel), tunnelling current (current channel), and friction (lateral force channel, LFM) across the sample surface. These measurements allow to visualize the moiré structure and the 2D fast Fourier transform of the moiré pattern served to determine the twist angle between the two graphene layers.

Nano-ARPES measurements were performed at the ANTARES beamline of the SOLEIL synchrotron (Saint-Aubin, France), equipped with a hemispherical analyzer (MB Scientific, type A-1) using a beam spot with a diameter of 600 nm. All measurements were carried out in ultra-high vacuum at a temperature of 70 K, using linearly horizontally polarized light with a photon energy of 100 eV. The energy resolution was 70 meV and the angular resolution 0.5° [31].

The simulations of TBG were performed using two different techniques – tight-binding (TB) method and DFT. The TB calculations were performed using an in-house Python implementation based on an atomistic π-orbital model, following the general approach of Ref. [32]. Each carbon atom was represented by a single $p_z$ orbital. The intralayer interactions were described by a nearest-neighbor hopping parameter of $t = -2.7$ eV, while the interlayer hopping was modeled using an exponentially decaying function of the interatomic distance, $t_\perp(d) = \gamma 1 \exp[-(d - d0)/\lambda]$, with $\gamma 1 = -0.35$ eV, $d0 = 3.35$ Å, and $\lambda = 0.45$ Å. Interlayer interactions were included up to a cutoff distance of 4.2 Å. To enable direct comparison with experiment, the calculated bands were unfolded onto the primitive Brillouin zone (BZ) of monolayer graphene. The unfolding procedure projects the supercell eigenstates onto primitive-cell Bloch states, yielding the spectral weight along the ($\overline{\Gamma} - \overline{\mathrm{M}} - \overline{\mathrm{K}} - \overline{\Gamma}$) high-symmetry path, following the formalism of Ref. [32].

The atomic supercells were generated using the supercell-core package [33], which searches for computationally tractable periodic supercells at a prescribed twist angle by optimizing the supercell geometry while minimizing the lattice strain required to accommodate the rotated layers. For the TB calculations, a supercell with a twist angle of 2.4°, containing 6892 carbon atoms (3446 atoms per layer), was selected. The supercell was defined by the matrix M = [[42, 41], [−41, 1]] and exhibited a maximum absolute strain-tensor component of approximately 0.018%, corresponding to a nearly unstrained structure. For the DFT calculations, a smaller supercell with a twist angle of 2.31° was selected to reduce the computational cost. The corresponding structure contained 390 carbon atoms (198 and 192 atoms in the bottom and top layers, respectively) and was defined by the matrix M = [[11, 6], [11, −3]]. The maximum absolute strain-tensor component was approximately 3.81%.

DFT calculations were carried out using the Quantum ESPRESSO simulation package [34] [35]. The electron-ion interactions were described via the projector-augmented wave method [36] [37]. We used the Perdew-Burke-Ernzerhof (PBE) parametrization of the generalized gradient approximation, and expanded the wave functions in a plane wave basis set with a cutoff of 60 Ry. A vacuum region of 17 Å along the c-axis was employed to separate the spurious images of bilayers, minimizing interactions between system replicas. Atomic positions were fully relaxed until the residual forces were below 10−4 Ry/Bohr, including van der Waals corrections via the DFT-D3 approach [38]. BZs were sampled on k-meshes of 3×3×1. The unfolding was performed using the BandUP(py) code [39-41].

## 3. Results and discussion

### 3.1. Structural characterization

TBG samples were prepared following the procedure described in the Methods section. The relative orientation of the two graphene layers was determined by conductive atomic force microscopy (c-AFM), which directly visualizes the moiré superlattice arising from the interference between the two graphene lattices. The c-AFM image in Figure 1 (a) presents a well-defined periodic moiré pattern extending over the investigated region, while its two-dimensional fast Fourier transform (Figure 1b) exhibits the expected sixfold symmetry. From the Fourier transform the wavelength of the moiré pattern, $\lambda_M$, was determined to be $5.8 \pm 0.2$ nm corresponding to a twist angle $\vartheta$ of $2.4 \pm 0.2°$ [42]. This value places the sample well above the first magic angle while remaining in the regime where moiré-induced modifications of the electronic structure are expected to be significant [8]. Importantly, no secondary periodicity associated with the *h*-BN substrate is observed in the c-AFM image. This confirms that the large rotational misalignment between graphene and the underlying *h*-BN effectively suppresses the formation of an additional graphene/*h*-BN moiré superlattice. Consequently, the electronic structure discussed below can be attributed predominantly to the TBG itself.

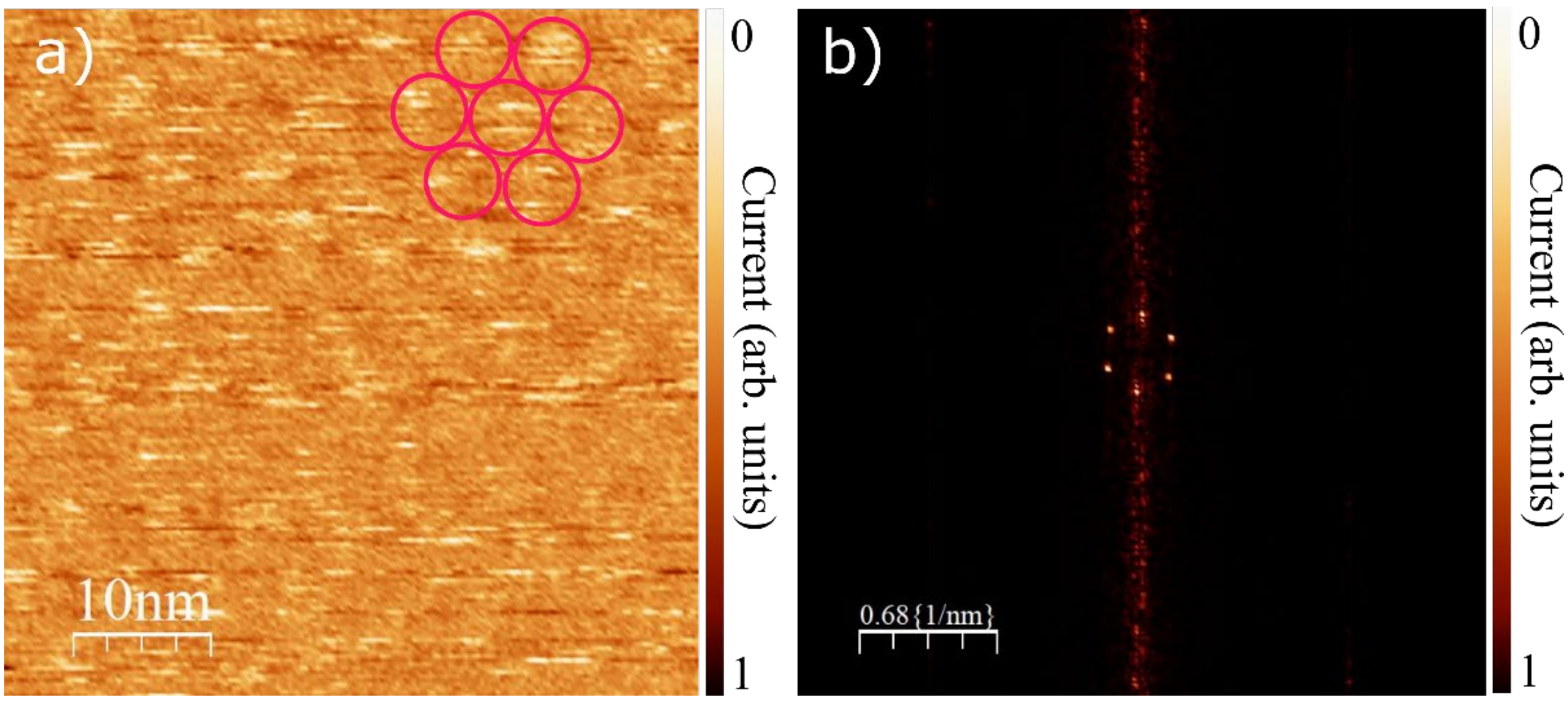

**Figure 1**: (a) Conductive atomic force microscopy (c-AFM) image showing a moiré pattern caused by the interference of the atom densities. (b) 2D fast Fourier transform of the moiré pattern exhibiting a hexagonal symmetry.

### 3.2. Electronic band structure measured by nano-ARPES

The electronic band structure measured by nano-ARPES is presented in Figure 2 along the high symmetry directions $\overline{\Gamma} - \overline{\mathrm{K}}$ (a) and $\overline{\Gamma} - \overline{\mathrm{M}}$ (b) of the Brillouin zone, additional data along the $\overline{\mathrm{A}} - \overline{\mathrm{K}} - \overline{\mathrm{A}}$ direction of symmetry can be found in Figure S1 in the Supplementary Material, together with the energy distribution curves that show the separation of the bands at the $\overline{\mathrm{K}}$ point (Figure S2). The spectra exhibit sharp and well-defined bands, demonstrating the high quality of the transferred structure and the suitability of the sample for momentum-resolved electronic structure measurements. Along the $\overline{\Gamma} - \overline{\mathrm{K}}$ direction two highly dispersive bands are resolved, corresponding to electronic states of the top and bottom graphene layers [43]. The valence band maximum (VBM) occurs at the $\overline{\mathrm{K}}$ point, in agreement with the theoretical predictions and existing literature for TBG [44]. A markedly different behavior is observed along the $\overline{\Gamma} - \overline{\mathrm{M}}$ direction (Figure 2b). While the two expected dispersive bands remain visible, [43] an additional band with substantially reduced dispersion appears close to the $\overline{\mathrm{M}}$ point at a binding energy of approximately 2.6 eV. The coexistence of this weakly dispersive feature with the ordinary graphene-derived bands immediately suggests an electronic state that cannot be explained by the conventional band structure alone [45]. A similar behavior is theoretically observed at the $\overline{\mathrm{M}}$ point in ref. [46] in bilayer graphene containing hexagonal vacancies, and in this case the flat dispersion is associated with the structural defects. However, due to the quality and sharpness of the bands at the $\overline{\mathrm{M}}$ and $\overline{\mathrm{K}}$ points, it is unlikely that these bands originate from vacancies [47, 48].

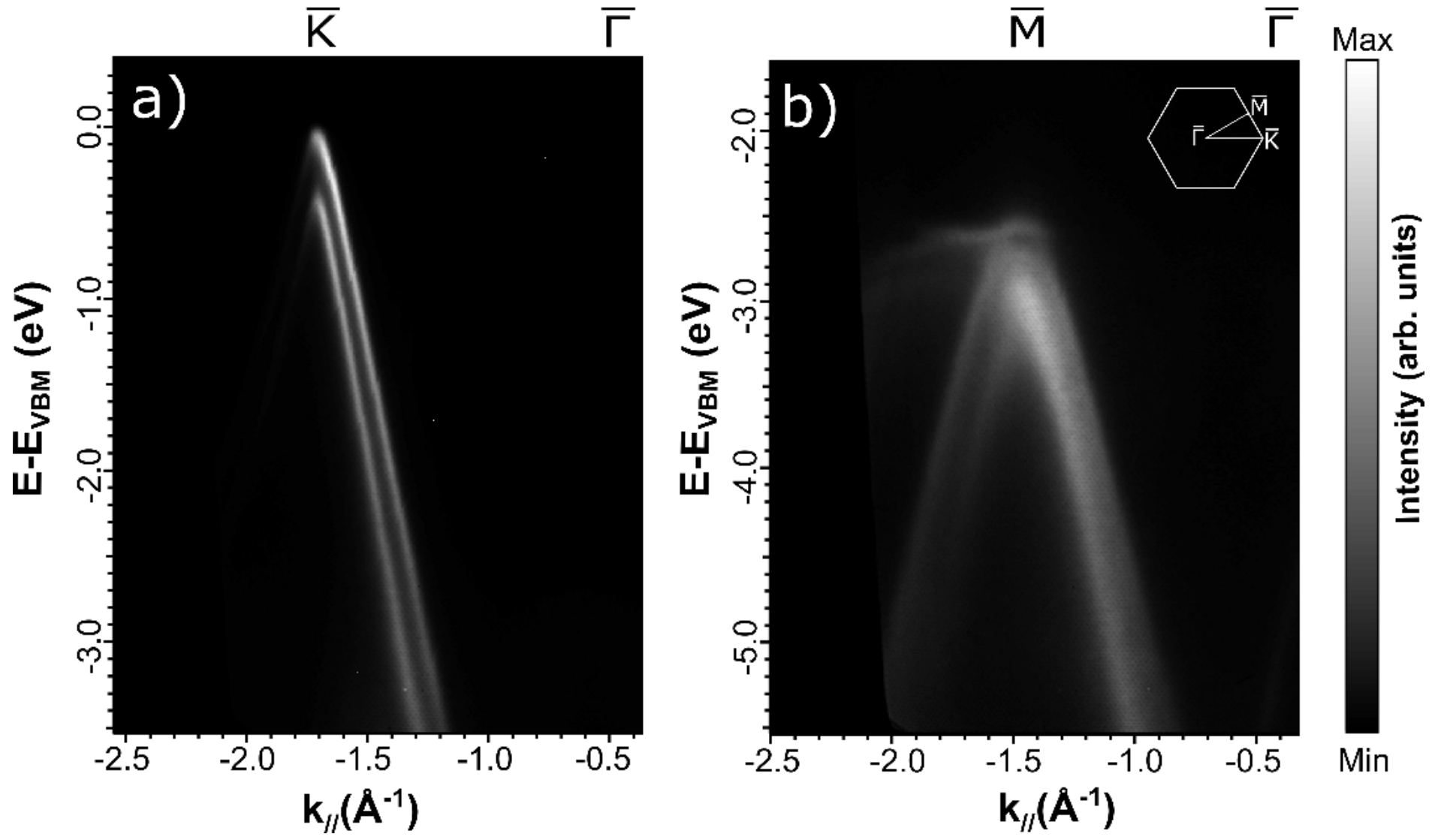

**Figure 2**: Energy band dispersion of twisted bilayer graphene (twist angle $2.4 \pm 0.2°$) along the two high symmetry directions $\overline{\Gamma} - \overline{\mathrm{K}}$ (a) and $\overline{\Gamma} - \overline{\mathrm{M}}$ (b). The high-symmetry directions are indicated in the Brillouin zone sketch in the inset in panel (b).

### 3.3. Curvature analysis reveals a flat band at the $\overline{\mathrm{M}}$ point

To examine the weakly dispersive band observed around the $\overline{\mathrm{M}}$ point in greater detail, we conducted a 2D curvature analysis of the nano-ARPES spectra. This procedure enhances weak dispersive features while preserving their momentum dependence, allowing closely spaced bands to be distinguished with improved clarity [49]. The results are shown in Figure 3 for both the $\overline{\Gamma} - \overline{\mathrm{K}}$ direction and $\overline{\Gamma} - \overline{\mathrm{M}}$ directions. Along the $\overline{\Gamma} - \overline{\mathrm{K}}$ direction (Figure 3a), only the two dispersive graphene bands remain visible, confirming the absence of any weakly dispersive states in the vicinity of the $\overline{\mathrm{K}}$ point and indicating possible weak interlayer coupling [17, 50]. In contrast, the 2D curvature image along the $\overline{\Gamma} - \overline{\mathrm{M}}$ direction (Figure 3b) clearly resolves an additional band with negligible dispersion (bandwidth 0.2 eV) around the $\overline{\mathrm{M}}$ point. The enhanced contrast provided by the curvature analysis demonstrates that this feature is not an artefact arising from overlapping dispersive bands but represents a distinct flat electronic state. Furthermore, at the $\overline{\mathrm{M}}$ point the Fermi velocity is $6.82 \times 10^4$ m/s, almost two orders of magnitude less than the Fermi velocity at the $\overline{\mathrm{K}}$ point, $1.14 \times 10^6$ m/s, which is comparable to that reported in literature for monolayer graphene [51], indicating weak interaction [8, 10].

The pronounced difference between the two high-symmetry directions highlights the strong momentum dependence of the electronic structure in the present TBG. While previous ARPES studies have focused primarily on flat bands in the vicinity of the $\overline{\mathrm{K}}$ point, our measurements reveal that weakly dispersive states may also emerge at other locations in reciprocal space. This observation underlines the importance of probing multiple momentum directions when investigating moiré-derived electronic states. To the best of our knowledge, this is the first observation of flat bands at a twist angle larger than the magic angle and away from the $\overline{\mathrm{K}}$ point.

Flat bands are typically observed only in the vicinity of the magic angle (1.1° for TBG) where interlayer coupling, and band hybridization strongly suppress the Fermi velocity. At larger twist angles, the increased separation of the Dirac cones generally weakens the hybridization, leading to more dispersive electronic bands [8]. The observation of a flat band under these conditions, and, moreover, near the $\overline{\mathrm{M}}$ point rather than the $\overline{\mathrm{K}}$ point, cannot be readily explained within the standard framework of magic-angle physics. Although the present measurements establish the existence of this flat electronic state, they do not by themselves identify its microscopic origin. To gain further insight, we compare the experimental electronic structure with atomistic calculations based on tight-binding and density functional theory.

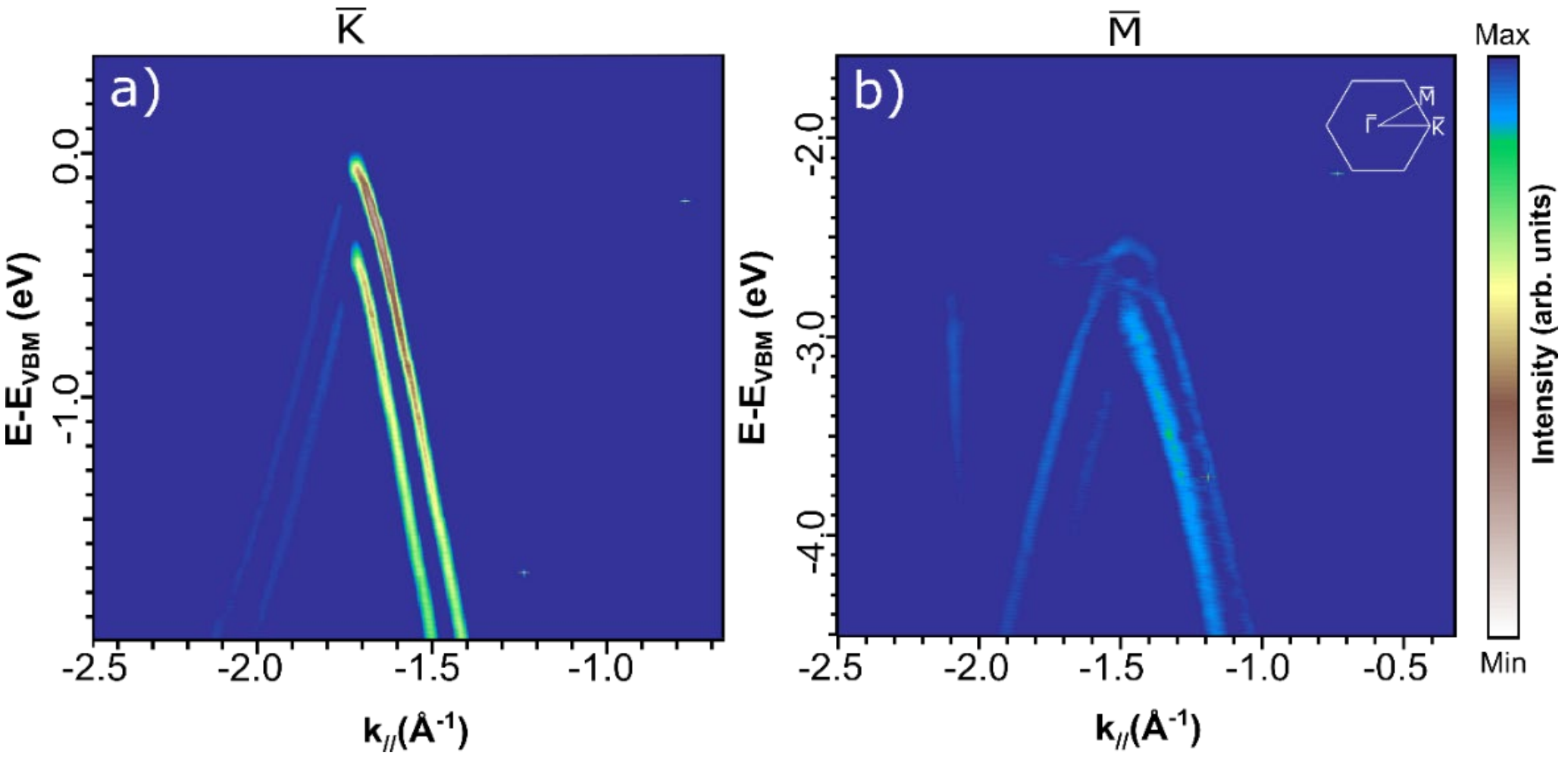

**Figure 3:** 2D curvature of the band structure of twisted bilayer graphene (twist angle $\mathbf{2.4} \pm \mathbf{0.2}°$) along the two high symmetry directions $\overline{\Gamma} - \overline{\mathrm{K}}$ (a) and $\overline{\Gamma} - \overline{\mathrm{M}}$ (b). 2D curvature data correspond to the energy dispersion spectra shown in Figure 2. Flat band dispersion is only present at the high symmetry point $\overline{\mathrm{M}}$, and completely absent at the $\overline{\mathbf{K}}$ point. High-symmetry points are indicated in the Brillouin zone sketch inset in (b)

### 3.4. Comparison with theoretical calculations

To understand the nature of the flat band at the $\overline{\mathrm{M}}$ point, we calculated the electronic structure of TBG performed within both the TB and DFT methods. Figure 4 shows the TB calculation performed for a twist angle $\vartheta$ = 2.4°, obtained using the supercell containing 6892 atoms. The calculation reproduces the principal features of the experimental electronic structure remarkably well. In particular, the dispersive bands around the $\overline{\mathrm{K}}$ point closely match the measured spectra, while the calculated band onsets near the $\overline{\mathrm{M}}$ point agree well with the experimentally observed dispersive states. This good overall agreement demonstrates that the electronic structure of the twisted bilayer is generally well described within the rigid-lattice approximation. However, despite reproducing the main dispersive bands, the TB calculation does not predict the additional flat band observed experimentally around the $\bar{M}$ point. This indicates that the weakly dispersive state cannot be explained solely by the electronic structure of an ideal rigid TBG. One possible explanation is that the rigid-lattice π-orbital-based approximation neglects structural effects, which also involve σ orbitals, known to occur in twisted van der Waals systems. Although lattice relaxation becomes progressively less pronounced with increasing twist angle, local structural distortions, residual strain, or domain boundaries may still modify the local electronic potential and give rise to electronic states that are absent in an idealized geometry. To explore this possibility, we performed complementary DFT calculations which include structural relaxation.

The DFT calculations, carried out for a bilayer with twist angle of 2.31°, likewise fail to reproduce the experimentally observed flat band (Figure S3 in Supplementary Material). Moreover, the overall agreement with experiment is less satisfactory than for the TB calculations.

This limitation primarily reflects the computational constraints imposed by the very large moiré supercell. The accessible supercell contains only 390 atoms, considerably fewer than required to faithfully describe the long-period moiré structure and possible local distortions associated with corrugation [52-54] and domain boundaries. Such domain walls can support quasi-one-dimensional electronic states [55-57], which may appear as flat or weakly dispersing bands in momentum-resolved measurements. Consequently, the calculations cannot fully capture long-range structural inhomogeneities that may influence the electronic band structure. The combined experimental and theoretical results therefore suggest that the observed flat band is not an intrinsic feature of an ideal periodic TBG. Instead, its appearance is consistent with an origin involving structural perturbations that are not incorporated in the present theoretical models. Such perturbations may include local strain, reconstruction, or other forms of symmetry breaking within the moiré lattice, including domain walls. At the same time, an additional contribution arising from residual interaction with the *h*-BN substrate cannot be completely excluded. Although the large graphene–*h*-BN twist angle suppresses the formation of an electronic moiré superlattice, weak hybridization effects at higher binding energies remain conceivable. Furthermore, one should also consider possible effects of strong electron-phonon coupling which is present across the range of angles in TBG [58], as electron phonon coupling has been known to induce kinks in monolayer graphene σ bands far from the Fermi level [59-62]. Distinguishing between these possible mechanisms will require calculations capable of treating realistic relaxed structures together with systematic experimental studies, for example combining spatially resolved structural characterization, electron energy loss spectroscopy [63] and nano-ARPES measurements on the same regions of the sample.

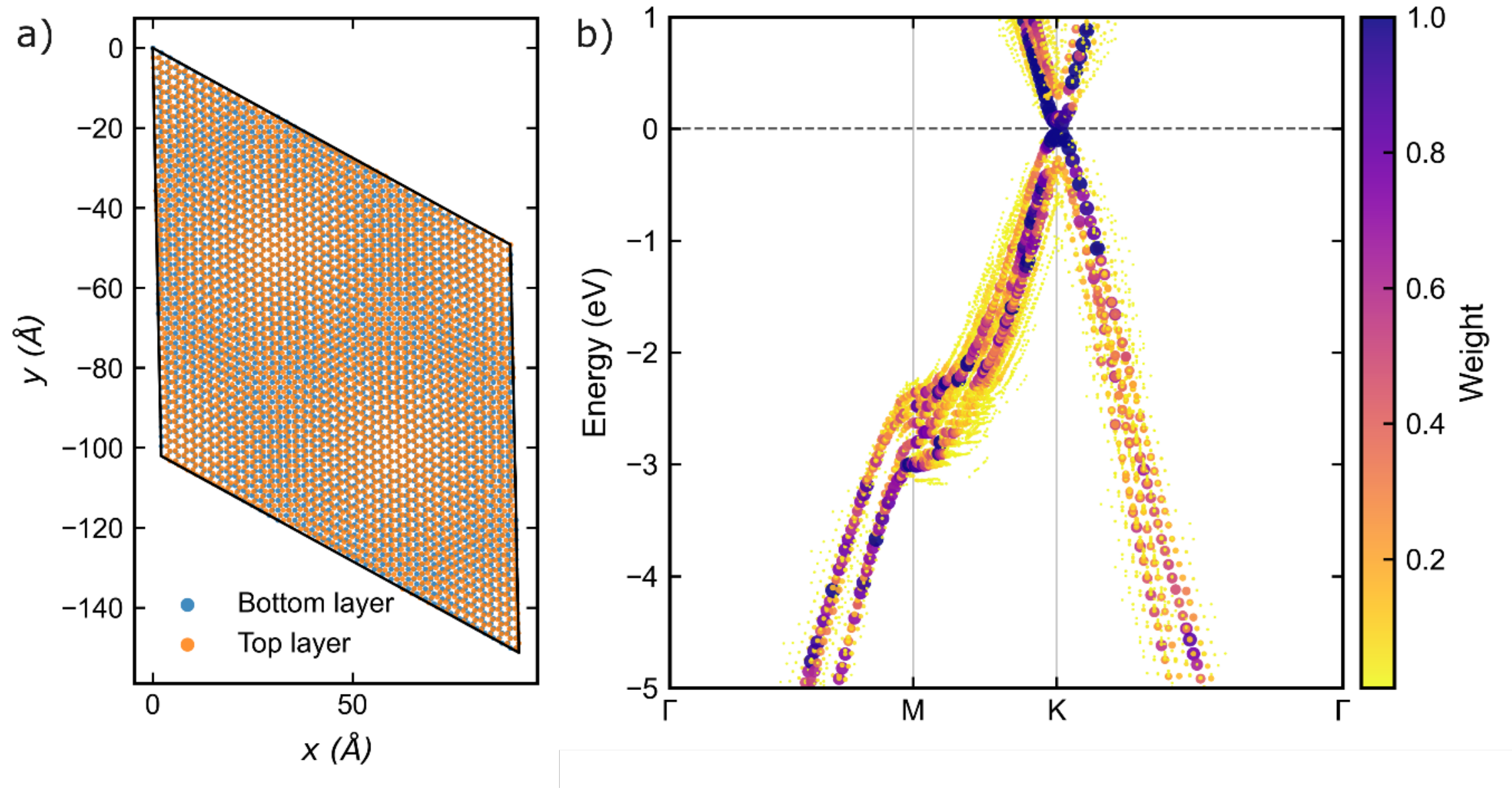

**Figure 4:** (a) The supercell of twisted bilayer graphene (twist angle 2.4°) used in the TB calculations. (b) TB calculations along the $\overline{\Gamma}-\overline{\mathrm{M}}-\overline{K}-\overline{\Gamma}$ direction. Different colors refer to different intensity of the bands.

## 4. Conclusions

In this work we investigated the electronic band structure of TBG with a twist angle of 2.4° by nano-ARPES, complemented by tight-binding and density functional theory calculations. The measurements reveal a pronounced weakly dispersing state around the $\overline{\mathrm{M}}$ point at a binding energy of approximately 2.6 eV, while no comparable feature was observed near the $\overline{\mathrm{K}}$ point. This distinguished the observed state from the flat bands associated with magic-angle TBG, which occur near the Fermi level and in the vicinity of the $\overline{\mathrm{K}}$ point.

The theoretical calculations successfully reproduce the principal dispersive features of the measured band structure but do not account for the additional flat band. This discrepancy indicates that the observed feature is unlikely to originate from the electronic structure of an ideal TBG and suggests that additional structural or interfacial effects, not captured by the present models, may contribute to its formation.

Our findings demonstrate that moiré-induced electronic structure changes in TBG are not necessarily confined to the low binding-energy region near the $\overline{\mathrm{K}}$ point. Instead, they point to a richer electronic landscape in which unexpected, localized states may emerge at larger twist angles, at different locations in the Brillouin zone, and at the higher binding energies. These results motivate further experimental and theoretical studies aimed at identifying the microscopic origin of such states and at developing more realistic models that incorporate structural relaxation, local inhomogeneity and substrate interaction.

## Acknowledgements

This work was financially supported by the National Science Centre, Poland under project 2018/30/E/ST5/00667 (P.P. and P.D.). The calculations were carried out on the Dutch national e-infrastructure with the support of SURF Cooperative (EINF-17379) and on the Hábrók high performance computing cluster of the University of Groningen. AGC acknowledges the financial support of the Zernike Institute for Advanced Materials.

Supplementary Material

for

# Nano-ARPES observation of a flat band at the $\overline{\mathrm{M}}$ point in 2.4° twisted bilayer graphene

Giovanna Feraco[1#], Przemysław Przybysz[1,2], Oreste De Luca[1$], Ying Wang[1%], Pavel Dudin[3], José Avila[3], Jianting Ye[1], Paweł Dąbrowski[2], Paweł J. Kowalczyk[2], Jagoda Sławińska[1], Petra Rudolf[1] and Antonija Grubišić-Čabo[1*]

[1]Zernike Institute for Advanced Materials, University of Groningen, Nijenborgh 3, 9747 AG Groningen, The Netherlands

[2]Faculty of Physics and Applied Informatics, University of Łódź, Pomorska 149/153, 90–236 Łódź, Poland

[3]Synchrotron SOLEIL, Université Paris-Saclay, Saint Aubin, BP 48, 91192 Gif sur Yvette, France

[#]Present address: European X-ray Free Electron Laser GmbH, Holzkoppel 4, 22869 Schenefeld, Germany

[$]Present address: Dipartimento di Fisica, Università della Calabria, Via Pietro Bucci, 87036 Rende, Italy.

[%]Present address: Department of Biomedical Signals and Systems, University of Twente, 7500 AE Enschede, The Netherlands

*Author to whom any correspondence should be addressed.

E-mail: a.grubisic-cabo@rug.nl

## *Additional ARPES data*

The electronic band structure measured by nano-ARPES along the $\overline{\mathrm{A}}-\overline{\mathrm{K}}-\overline{\mathrm{A}}$ direction of symmetry can be found in Figure S1, while with the energy distribution curves that show the separation of the bands at the $\overline{\mathrm{K}}$ point are presented in Figure S2.

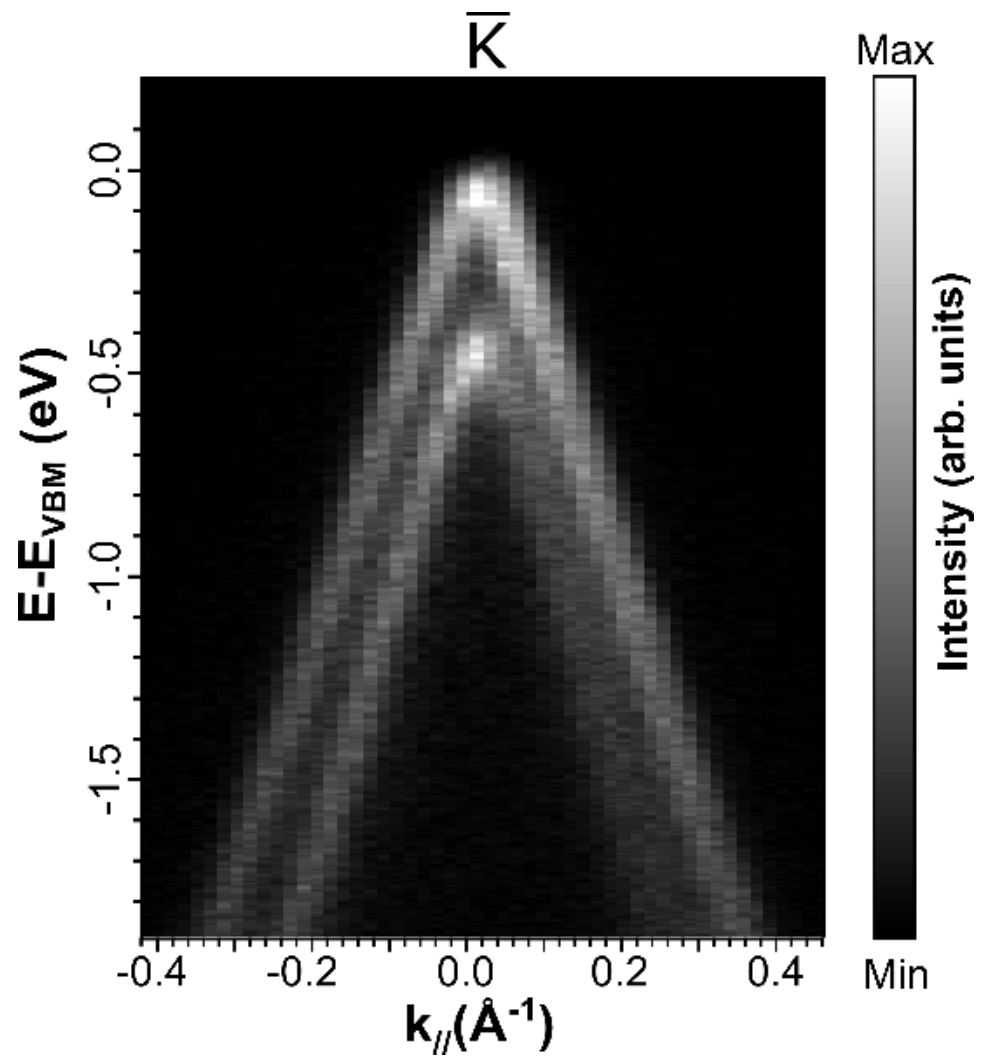

*Figure 1: Nano-ARPES along the $\bar{A}-\bar{K}-\bar{A}$ symmetry direction; the data show no flat band.*

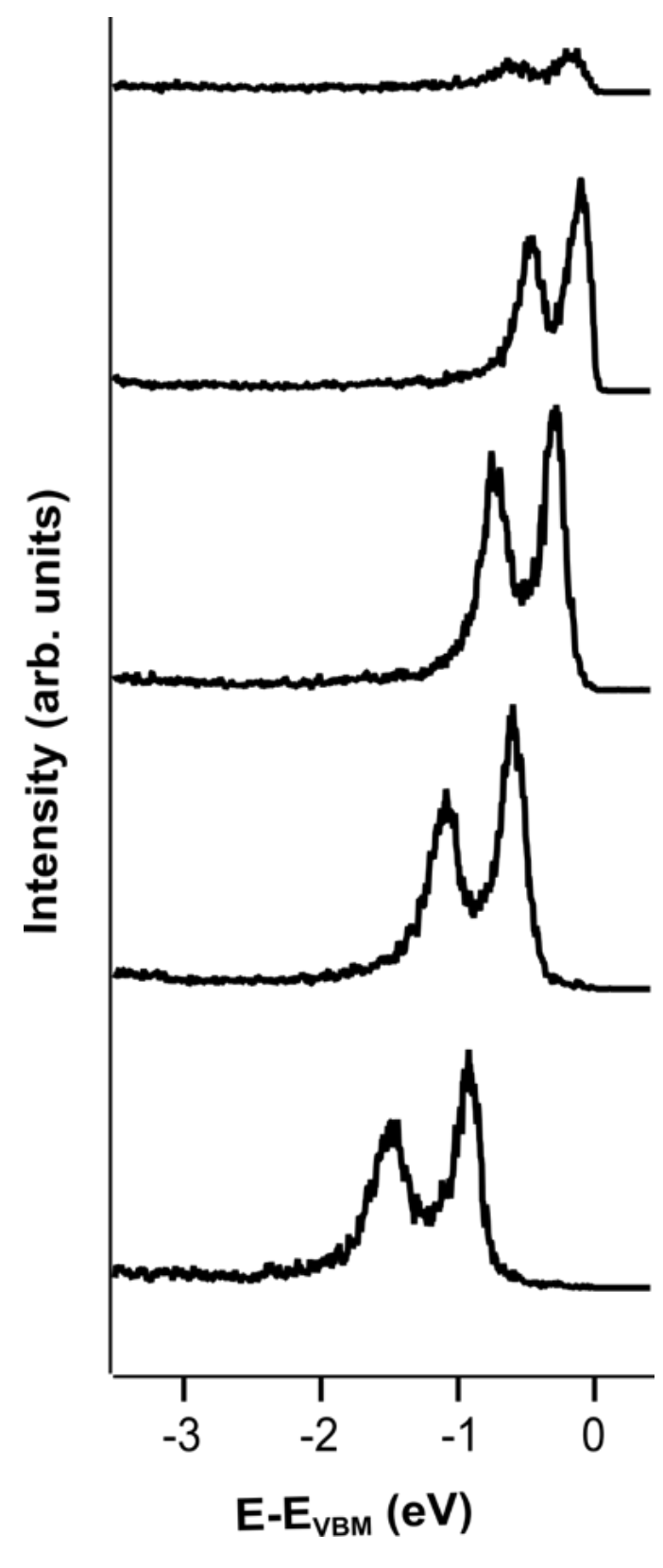

*Figure 2: Energy distribution curves that show the separation of bands at the $\bar{K}$ point.*

## *Additional theoretical data*

The DFT calculations, carried out for a bilayer with twist angle of 2.31°, do not show the experimentally observed flat band.

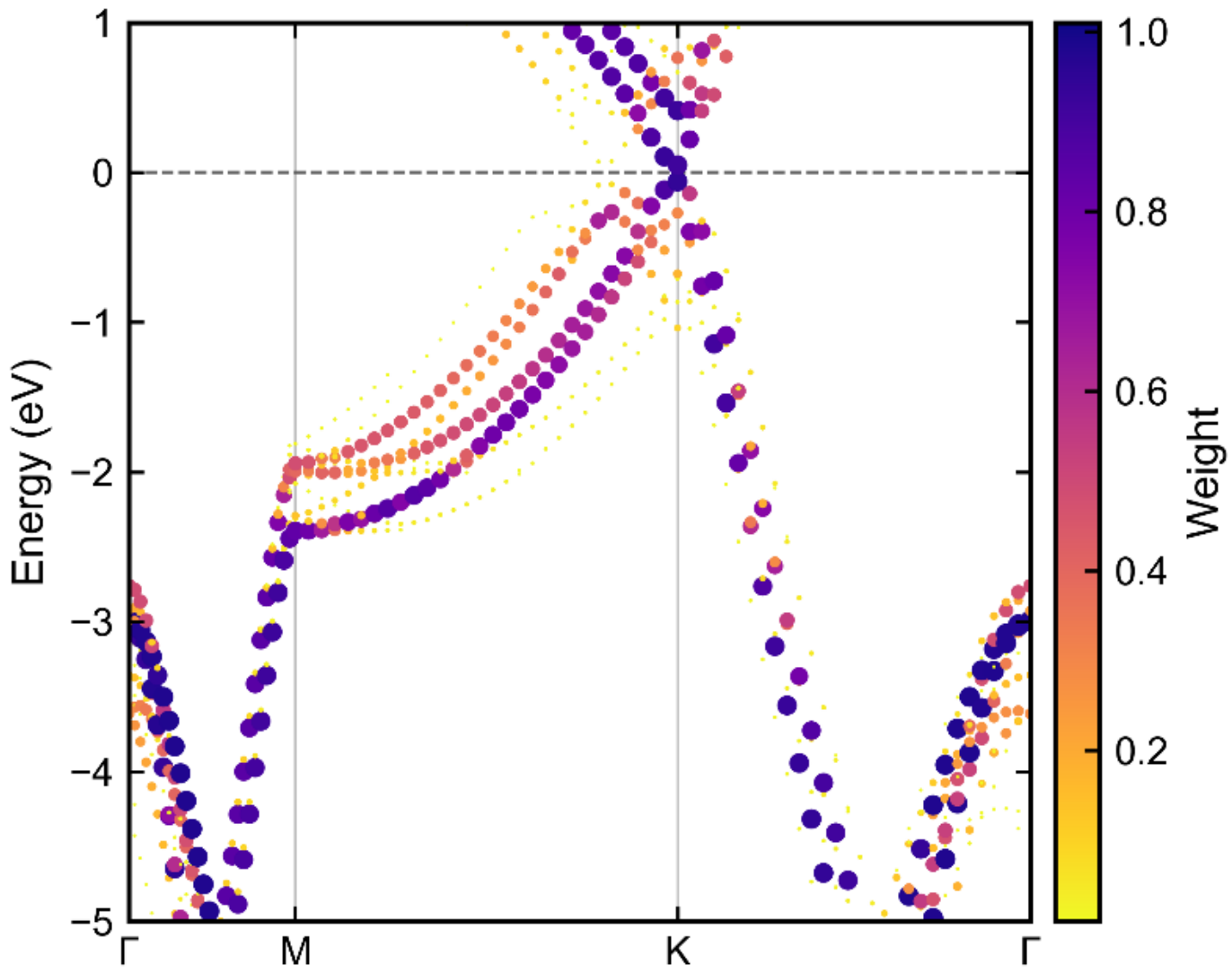

*Figure 3: DFT calculations of the band structure of bilayer graphene with a twist angle $\vartheta$ = 2.31°. The data show no flat band.*